# Size-independent Shear Band Formation in Amorphous Nanowires Made from Simulated Casting

Yunfeng Shi

Department of Materials Science and Engineering,

Rensselaer Polytechnic Institute, Troy, New York 12180, USA

**Abstract**

Molecular dynamics simulations indicate that surfaces strongly influence the strain localization behavior of amorphous nanowires in tension. A sample preparation routine that simulates casting was employed to facilitate the relaxation of the sample surface. Samples as short as 15 nm (7.5 nm in diameter) form dominant shear bands during deformation. The elastic energy release during plastic deformation is sufficient to provide the excess potential energy required for the shear band nucleation at rather small sample sizes. The results show that shear band formation is almost size-independent and is bounded only by its own length scale.

Strain localization is an important deformation mode for numerous material systems including metallic glasses (MGs).[1] The propensity of strain localization in a MG sample strongly affects its plastic response and hence the usefulness in load-bearing applications. The emergence of shear bands is likely due to structural softening instead of thermal softening. It has been shown both numerically[2] and experimentally[3] that the strain localization behaviors are sensitive to the initial structure of the testing sample, which can be tuned through processing. It was further demonstrated that such structural softening is manifested as the disordering of certain short-range structural signatures.[2,4,5]

In addition to the atomic structure of the sample, shear band formation may also be influenced by extrinsic factors such as loading conditions,[6] microstructures[7] and the sample size.[8-16] Particularly, the size-dependent plasticity in MG systems has attracted considerable attention due to the advancement in micromechanical testing and sample preparation using focused-ion beam technique. Two related size-dependent plasticity have been investigated. The first size-dependent phenomenon is the maturation and propagation of a shear band to a crack. It has been shown that the shear band grows stably for samples with a dimension of about 100 *nm* to large tensile strain instead of the typical brittle failure for macroscopic samples,[8] along with other demonstrations.[9-12] The second size-dependent effect concerns the deformation mode during the early stage of plasticity. A deformation mode transition from localized to homogeneous deformation was reported as the sample diameter is reduced below 400 *nm*.[12] Similarly, significant plastic deformation in samples with 330 *nm* in diameter was observed to be carried out in homogenous flow with or without the occurrence of a major shear band.[9] However, to the contrary, serrated flow was found in Pd-based MG samples with sizes ranging from 20 *μm* to 250 *nm*.[13] The same size-independent deformation behavior was later confirmed in Zr-based MG systems[14,15] with a diameter as small as 150 *nm*.

Molecular level simulation techniques are ideal to study size-dependent plasticity, as demonstrated in nanocrystalline materials [17] and crystalline nanowires,[18] and provide unparalleled opportunities to control the preparation and testing of nanoscale samples. Particularly, a perfect cylindrical sample can be made without a taper geometry or implanted impurities. However, the deformation modes of nanoscale MG samples observed in molecular simulations have also been inconsistent. A highly inhomogeneous deformation mode resulting in a major shear band has been observed in two-dimensional[4,19] and three-dimensional thin-slab samples.[5,20,21] Nonetheless, cylindrical-shaped samples, which are most relevant to experiments, have shown nearly homogeneous flow until necking at high strains [22,23] and only exhibit shear bands upon the introduction of a surface notch.[24,25] The central question is whether the emergence of a shear band is a result of special dimensionality or sample geometry, and whether the absence of shear bands in cylindrical samples is due to the small sample size. One important issue that has so far been overlooked is the sample preparation procedure, as previous cylindrical-shaped samples were usually obtained by cutting from a bulk sample at a low temperature.[22,23] Thus-created surfaces are highly energetic and unlikely to undergo sufficient relaxation prior to mechanical tests. Such surface effects are more pronounced for cylindrical samples in three dimensions.

Here, we employ a new sample preparation method to create cylindrical glassy samples by cooling a liquid confined within cylindrical walls, termed "simulated casting". This procedure allows sufficient surface relaxation during the entire quenching process as opposed to limited relaxation upon creating surfaces at low temperatures in the traditional cutting procedure. We choose to simulate a binary Lennard-Jones (LJ) glass-forming system with force field parameters devised by Wahnstrom.[26] The low temperature mechanical behavior has been systematically studied in uniaxial compression[5] and nanoindentation[2] both in thin-slab geometry. The closest binary MG system is $Ni_{50}Nb_{50}$ in terms of composition and atomic radius ratio. All physical quantities will therefore be expressed in SI units following the conversion in a previous report.[2]

The cylindrical-shaped amorphous solid samples are prepared using simulated casting as follows. The initial system is a high temperature liquid confined within a cylindrical repulsive wall. The periodic boundary condition is applied only along the cylindrical axis direction. There is repulsion between the atom and the wall that follows $k(r-r_0)^2/2$ if $r$ (the distance of the atom to the cylindrical axis) is greater than $r_0$ (the radius of the wall). This interaction is zero if if $r$ is less than $r_0$. The initial spring constant $k$ is 11.1 $kg/s^2$. We follow the quenching protocol of WA-3 samples in Ref. [5], which have been shown to exhibit shear band formation. In addition, the spring constant $k$ of the atom-wall interaction decreases linearly to zero during cooling to ensure the final sample is stress-free. Using simulated casing, four series of samples with five independent samples were prepared with the same diameter-to-length aspect ratio (1:2) and a length ranging from 5 to 20 $nm$. Additional samples were also prepared via bulk cutting. The samples from cutting were relaxed for 0.1 $ns$ before mechanical testing. The radial potential energy profile shows that the interior materials from both preparation methods have identical potential energy, while the surface layer in samples from casting has a slightly lower potential energy than those from the cutting method (Figure S1). The surface of the samples from casting contains more atoms belonging to the larger species likely due to their lower mobility. By imposing species-dependent atom-wall interaction, preliminary results show that such composition does not affect the deformation mode.

Uniaxial tensile tests were conducted for the cylindrical-shaped samples with a strain rate of 100 $\mu s^{-1}$. Typical stress-strain curves of samples with different sizes are shown in Figure 1. The Young's modulus for the samples from casting as shown ranges between 64 to 67 $GPa$ (Figure S2). The stress-strain curve of the smallest sample is considerably rougher than larger samples due to the large relative stress fluctuations induced by individual plastic events. The first plastic event, as indicated by the stress-drop in the 5 nm sample, occurs at about 2.0% strain. The tensile strength of the smallest sample is the highest and those of the rest samples seem to decrease slightly with the growing sample size. Figure 1 also shows that the sample from cutting is 10% lower in Young's modulus and tensile strength than the sample from casting at the same size.

Regions of plastic deformation are identified by calculating the local deviatoric shear strain. Inset graphs in Figure 1 show the deformation morphologies for different samples at 10% tensile strain (complete deformation morphologies shown in Figure S3 and animations in Supplemental Material). Large samples deform via shear band formation, while small samples exhibit necking instead of shear slip along an inclined plane. The sample created by cutting appears to deform homogeneously until it forms a neck, which resembles previous numerical results.[22,23] Thus, it is conceivable that previous reports of the homogeneous flow of cylindrical-shaped model MGs might be due to the sample preparation procedure instead of the sample size. Similar caution might also be applicable to samples fabricated by focused ion beam which contain structural damages and chemical impurities.[12] The fact that the deformation of nanoscale samples is quite sensitive to their surfaces calls for careful sample surface treatment in simulations and experiments.

Both slipping along a shear band and necking without a shear band are localized. The difference is the orientation of the deformed material. The normal direction of the deformed plane can be calculated by minimizing the sum of the projected distances for all deformed atoms to the deformed plane anchored by its center of mass. Figure 2 shows the angle between the normal of the deformed plane and the loading direction. For large samples, the angle is around 35 degrees which indicates the contribution of normal stress to yielding in addition to shear. For small samples, the angle is considerably smaller. It is also clear that samples from cutting of all sizes deform via necking with a small angle. Thus, the deformation mode changes from slip along a shear band to necking at a sample length of 10 *nm*.

We examined the emergence of shear band inside nanowire samples during tension from an energetic perspective. Similar analysis has been reported by Volkert, Donohue and Spaepen.[12] In their analysis, a critical stress value was identified by assuming a complete release of elastic energy and 100% plastic strain inside the shear band. Here, we recognized from Figure 1 that only a portion of the elastic energy is released during the shear band formation. Moreover, the potential energy of the shear band can be directly measured in simulations. Figure 3 shows the average potential energy as a function of relative distance away from the shear plane for the largest sample at different strain. The mechanical disordering appears around 6% strain and continues to grow thereafter. The steepest drop as shown in the stress-strain curve occurs between 7 to 8% strain. Thus the shear band initiates around 6% strain and develops across the sample around 8%. The half-height width of the potential energy profile at 8% strain is about 2 *nm* (thus a 4 *nm* shear band thickness) and increases as deformation proceeds. Extending the analysis by Volkert, Donohue and Spaepen,[12] the partial elastic energy release is $\Delta E^{elastic} = \pi r^2 l \left( \sigma_Y^2 - \sigma_F^2 \right) / 2E$, where $E$ is the Young's modulus, $r$ is the radius, $l$ is the length of the sample, $\sigma^Y$ and $\sigma^F$ are the yielding and flow stress, respectively. The shear band energy is $\Delta E^{shearband} = \pi r^2 t \cdot n \cdot \Delta w / \cos\theta$, where $t$ is the half height width and $\Delta w$ is the peak height of the potential energy profile of the shear band, $\theta$ is the angle between the shear plane normal and the loading direction, $n$ is the number density. The necessary condition is that the elastic energy release has to be greater than the shear band energy. Thus, $l_{min} = \dfrac{E \cdot n \Delta w}{\left( \sigma_Y^2 - \sigma_F^2 \right)} \dfrac{2}{\cos\theta} t$.

Typical parameters measured in simulation are $t=2$ *nm*, $n= 6.5\times10^{28}$ $m^{-3}$, $E=67$ *GPa*, $\Delta w=0.01$ *eV*/atom, $\sigma^Y=2.1$ *GPa*, $\sigma^F =1.4$ *GPa*, $\theta=30°$. The threshold length is estimated to be about 13 *nm*. It should be noted that such consideration is quite crude for (1) the omission of heat generation from elastic energy release and (2) the possibility of partial shear band formation. Nonetheless, the estimation of 13 *nm* is close to the transition length of 10 *nm* as shown in Figure 2.

The deformation mode during early-stage plasticity is via slip along shear bands unless the sample length is close to the shear band thickness, which appears to be an almost size-independent behavior. Such behavior is not unexpected from the fact that amorphous solids are structureless beyond a few atomic spacings. The only relevant length scale is probably the thickness of the shear band itself, which seems to be the spatial limitation in the current system. It should be noted that the catastrophic crack propagation, which is indeed strongly size-dependent, is likely to be a thermo-mechanical process that may inherit new length scales from thermal diffusion in addition to the intrinsic structural length scale of MG samples.[1,27]

In summary, we have prepared amorphous samples in cylindrical shapes with a new simulated casting procedure with a relaxed surface layer. During uniaxial tensile tests, major shear bands develop in samples as short as 15 *nm* in length, which is only 3-4 times the shear band thickness. The simulation results here demonstrate that amorphous samples exhibit shear banding which appears to be size-independent until approaching the shear band thickness. It is further shown that the elastic energy release is sufficiently large to provide the excessive energy associated with the shear band region.

We thank Michael Falk for careful proof-reading of the manuscript and encouragements. We are also in debt to Lallit Anand, Todd Hufnagel, Craig Maloney and Liping Huang for stimulating discussions.

**Figure Captions**

Figure 1. Stress-strain curves for samples from casting with lengths from 5 to 20 nm (deformation morphologies at 10% strain are shown as insets). The sample from cutting with a length of 20 nm at the same strain is also shown (far left). The samples are rotated along the cylinder axis such that the shear plane is perpendicular to the paper plane. The atoms are colored from red (0% shear strain) to yellow (20% or larger shear strain).

Figure 2. The angle between the shear plane normal and the loading direction is used to differentiate shear banding and necking (insets). The angles are plotted as a function of the sample length for both samples from casting (circles with error bars from five independent simulations) and samples from cutting (filled circles).

Figure 3. The potential energy profiles along the relative distance to the shear plane for one of the largest samples from casting for strains from 5% to 10%.

Figure 1

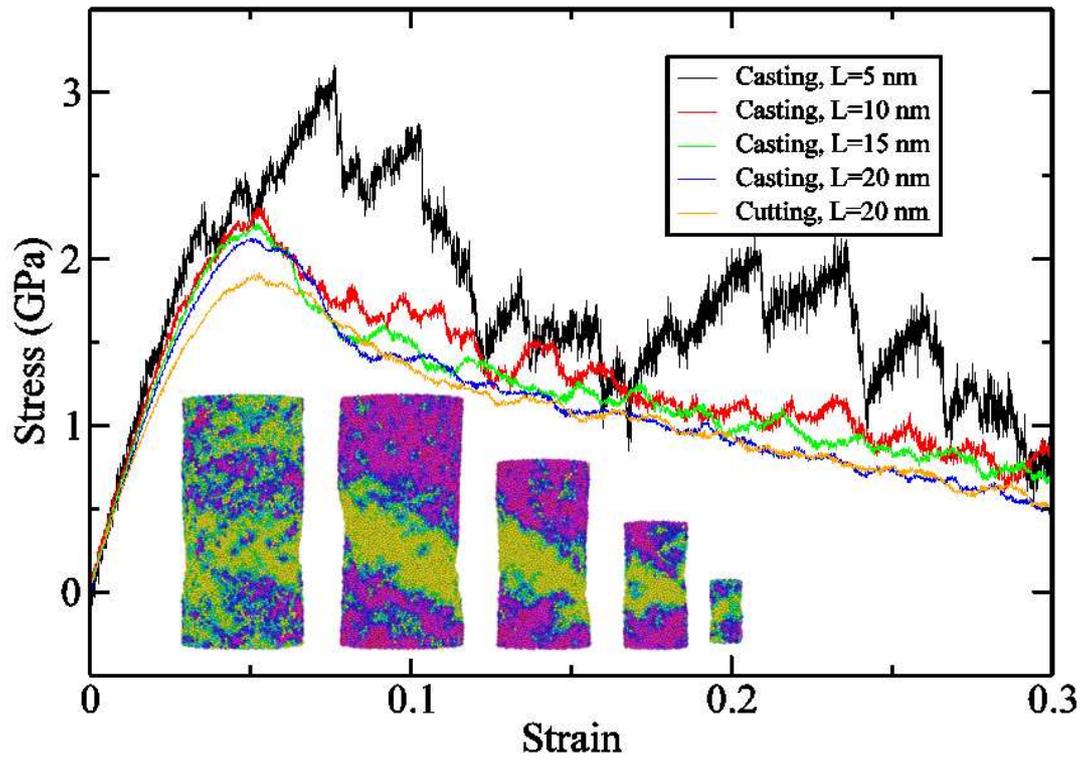

Figure 2

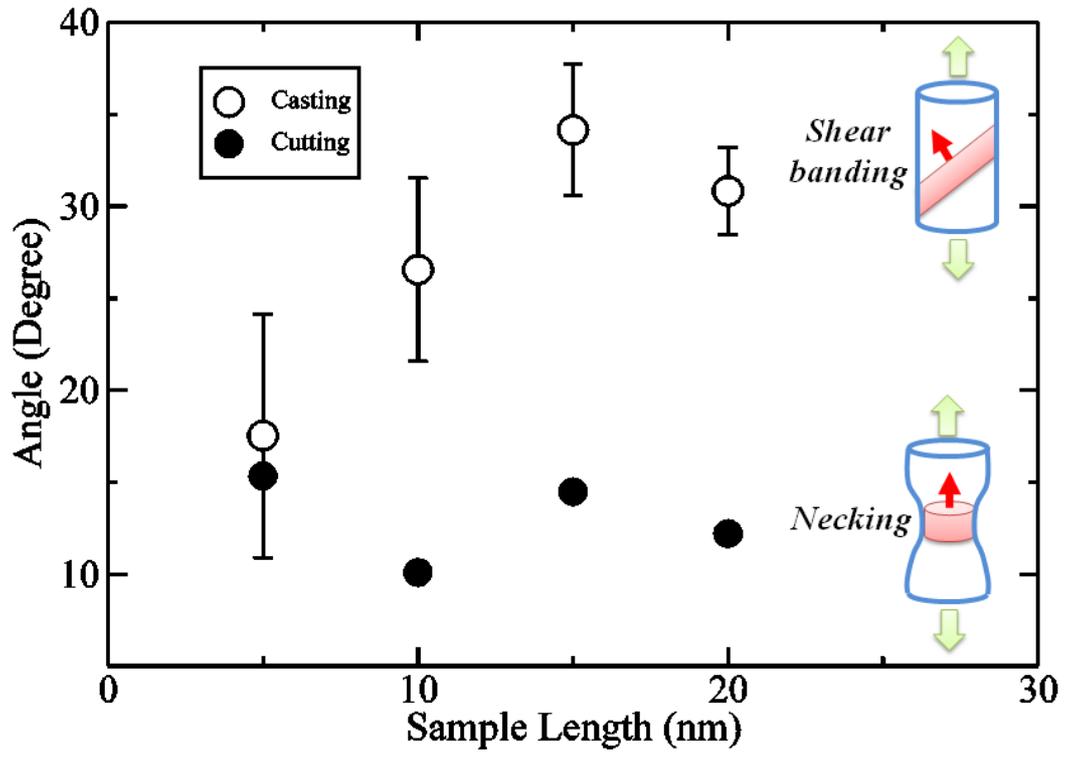

Figure 3

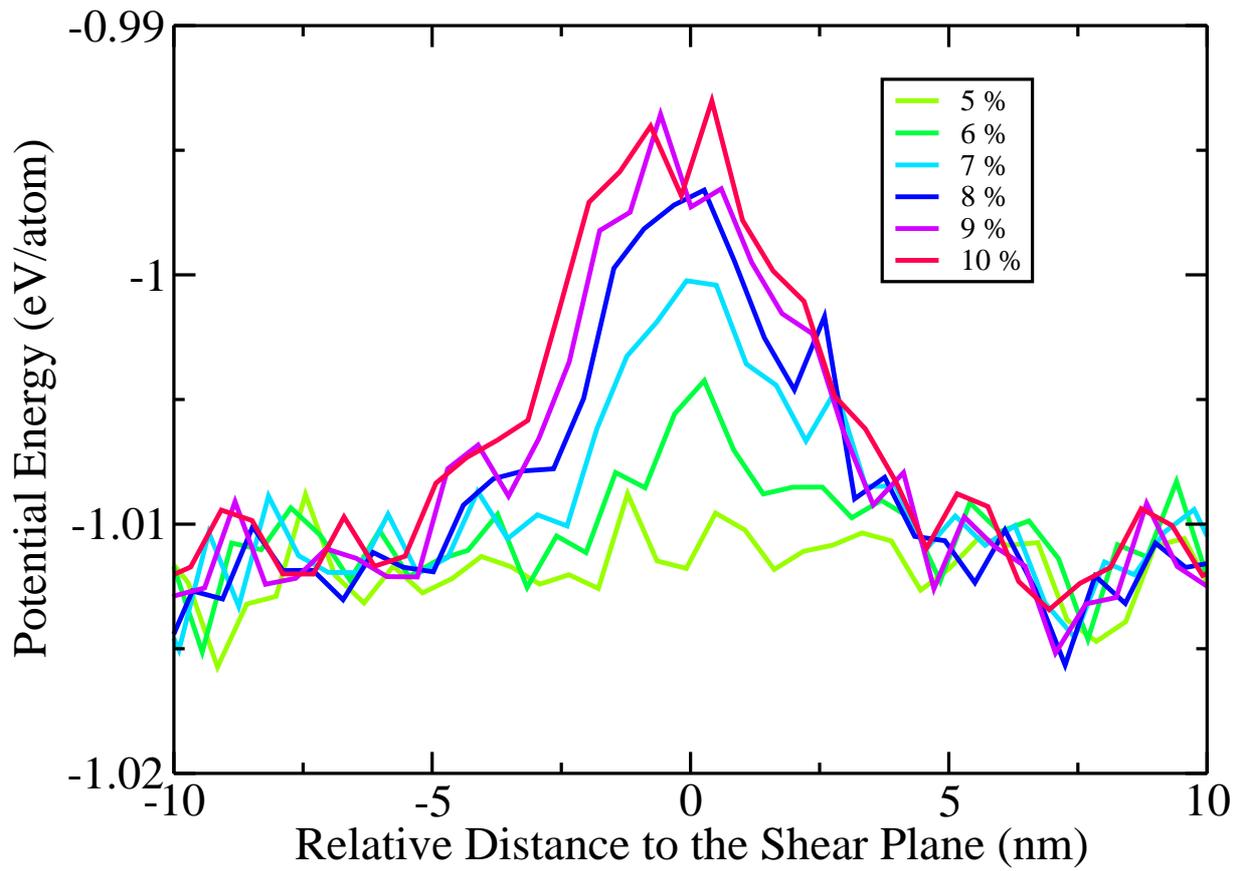